\documentclass{JHEP3}

\usepackage{amsmath,graphicx} 

\newcommand{\centeron}[2]{{\setbox0=\hbox{#1}\setbox1=\hbox{#2}\ifdim
\wd1>\wd0\kern.5\wd1\kern-.5\wd0\fi \copy0
\kern-.5\wd0\kern-.5\wd1\copy1\ifdim\wd0>\wd1
                                    \kern.5\wd0\kern-.5\wd1\fi}}
\newcommand{\ltap}{\>\centeron{\raise.35ex\hbox{$<$}}
                            {\lower.65ex\hbox{$\sim$}}\>}
\newcommand{\gtap}{\>\centeron{\raise.35ex\hbox{$>$}}
                            {\lower.65ex\hbox{$\sim$}}\>}

\newcommand\ZZ{\hbox{\zfont Z\kern-.4emZ}}
\font\zfont = cmss10 

\newcommand{\eref}[1]{eq.\ (\ref{e.#1})}

\newcommand{\cref}[1]{Chapter \ref{c.#1}}

\def\nn{\nonumber \\}

\def\beq{\begin{equation}}
\def\eeq{\end{equation}}
\newcommand{\ba}{\begin{array}}
\newcommand{\ea}{\end{array}}
\newcommand{\bea}{\begin{eqnarray}}
\newcommand{\eea}{\end{eqnarray} }
\newcommand{\bal}{\begin{align}}
\newcommand{\eal}{\end{align}}
\def\bi{\begin{itemize}}
\def\ei{\end{itemize}}
\def\ben{\begin{enumerate}}
\def\een{\end{enumerate}}
\def\beq{\begin{equation}}
\def\eeq{\end{equation}}
\def\bc{\begin{center}}
\def\ec{\end{center}}
\def\bt{\begin{table}}
\def\et{\end{table}}
\def\btb{\begin{tabular}}
\def\etb{\end{tabular}}

\def\co{{\mathcal O}}

\def\mass2{mass${}^2$}

\def\ra{\rangle}
\def\la{\langle}

\def\eps{\epsilon}

\title{Unparticle signals with a few particles}

\author{Manuel P\'erez-Victoria\\ CAFPE and Departamento de F\'{\i}sica Te\'orica y del Cosmos, \\ Universidad de Granada, E-18071, Spain \\ E-mail: \email{mpv@ugr.es}}

\abstract{We use Pad\'e approximants to systematically approximate scalar unparticle propagators and their associated phase factors by a finite number of ordinary particles. This is possible for conformal dimensions $1\leq d<2$, and also for $d\geq 2$ if we add local terms. A small number of particles can, in some cases, mimic unparticle signals. We also discuss how the approximants capture basic unparticle properties.}


\preprint{arXiv:0808.4075 [hep-ph]}

\begin{document}

\section{Introduction}
Unparticle physics \cite{G1} refers to the phenomenology of a conformally invariant sector (CFT) coupled to the Standard Model (SM). The main motivation for studying these scenarios at the dawn of the LHC era is that unparticles can give rise to novel signals, both in production and in virtual exchanges~\cite{G1,G2,Cheung:2007zza,Cheung:2007ap}. Here, we argue that these effects can be reproduced by a few particles with specific masses and couplings.

That the existence of such a description is plausible can be seen right away from the following two observations. First, a scalar operator with the smallest conformal dimension allowed by unitarity describes a free particle. Continuity in the dimension then implies that unparticle stuff with non-integer dimension close to the unitarity bound can also be well approximated by one ordinary particle. On the other hand, it has been shown~\cite{Stephanov} that the propagators of unparticles arise as a continuum limit of an infinite discrete set of particles with a particular pattern of masses and residues. Therefore, it seems not unreasonable to interpolate between the one-particle and infinite-particle approximations and try with a finite number of particles. We shall make this idea explicit with the help of Pad\'e approximants. This is inspired by Migdal's proposal of reconstructing low-energy QCD large-$N_c$ correlators from their UV value by means of Pad\'e approximants and a particular double limit~\cite{Migdal}. We stress, nevertheless, that we do not take that limit here.

We concentrate on two-point functions, for simplicity, even though the CFT interactions can be very relevant for collider phenomenology~\cite{Strassler,Feng}. Moreover, we study scalar operators only. We consider first exactly conformal unparticles\footnote{Note however that the coupling to the SM breaks conformal invariance to some degree, mostly through the vev of the Higgs~\cite{Fox,DEQ}.} and then discuss the case of unparticles with a mass gap. 
Because the functions we want to approximate have a branch cut for timelike Minkowskian momentum, we choose the Pad\'e point (at which the approximation is best) at an arbitary Euclidean momentum, that we identify with the UV renormalization scale.

The paper is organized as follows. In Section~2 we review basic properties of conformal two-point functions (unparticle propagators), and find the corresponding Pad\'e approximants. In Section~3 we use these formal results to approximate unparticle stuff by a finite set of particles. We argue that these particles can mimic characteristic unparticle signals, and give explicit examples of this. In Section~4 we study how the Pad\'e approximants capture fundamental unparticle properties related to the value of the conformal dimension. Section~5 is devoted to the case of unparticles with a mass gap. We conclude in Section~6.

\section{Pad\'e approximants of unparticle propagators}
Let us start reviewing some simple facts that will be important below. We use a mostly minus Minkowski metric. Primary scalar operators have a conformal dimension $d\geq 1$, with $d=1$ in a free theory. Their two-point function in position space is 
\beq
\la \co_d(x) \co_d(0) \ra = C_d \frac{1}{(x^2)^d}.
\eeq
This is a {\it bona fide\/} distribution for $d<2$. When $d\geq 2$, however, it is too singular at $x=0$. The divergent local terms can be substracted to give a finite renormalized Fourier transform,
\beq
\la \co_d(p) \co_d(-p) \ra_\mathrm{ren} = A_d \frac{\pi (\nu+1)}{\sin (\pi \nu)} (-t)^\nu + P_\nu(t) \equiv \Delta_d(t),
\eeq
We have defined $\nu=d-2$ and $t=p^2/\mu^2$, with $\mu$ the renormalization scale. Similarly, all dimensional quantities that appear below are made dimensionless by writing them in units of $\mu$.
For $\nu\geq 0$, $P_\nu$ is a polynomial of degree $[\nu]$ (where $[\nu]$ is the greatest integer not larger than $\nu$) while for negative $\nu$, $P_\nu=0$. The relevance of the local terms $P_\nu$ has been emphasized and discussed in~\cite{GIR} (see also~\cite{MPV,Terning} for their AdS/CFT dual interpretation). In the following, we assume a particular substraction scheme and fix $P_\nu$ such that the first $[\nu]+1$ coefficients in the Taylor series of $\Delta_d(t)$ about $t=-1$ vanish. We also normalize to $A_d=1$.

The renormalized two-point function $\Delta_d(t)$ is the unparticle propagator. For the discussion below, let us also distinguish its non-local part $\tilde{\Delta}_d=\Delta_d-P_\nu$. When $1<d<2$, the propagator can be written in a K\"ahlen-Leman spectral representation
\beq
\Delta_d(t) = \int_0^\infty \frac{\sigma_d(w)}{t-w}\, \mathrm{d}w , \label{e.spectral}
\eeq
with a spectral density
\begin{align}
\sigma_d(w) &= -\frac{1}{\pi} \mathrm{Im} \Delta(w+i \eps) \nn
& = (\nu+1) w^\nu.
\end{align}
Note that $\sigma_d(w)$ gives also the phase factor for unparticle production~\cite{G1}.
For $d\geq 2$ the integral diverges, but we can use instead a substracted dispersion relation:
\beq
\Delta_d(t) = (t+1)^{[\nu]+1} \int_0^\infty \frac{\sigma_d(w)}{(w+1)^{[\nu]+1}}\frac{1}{(t-w)}\, \mathrm{d}w .
\eeq
Changing variables to $u=(w+1)^{-1}$, it reads
\beq
\Delta_d(s-1) =  -s^{[\nu]+1} \int_0^1 \frac{W(u)}{1-s u} \mathrm{d}u ,
\eeq
where $W(u)=u^{[\nu]}\sigma_d(u^{-1}-1)$ and we have introduced the variable $s=t+1$. By definition, this means that $\Delta_d(s-1)/s^{[\nu]+1}$ is a Stieltjes function of $s$ with measure $W(u) \mathrm{d}u$.

Our plan is to study the Pad\'e approximants of these unparticle propagators. Let us first review the definition of Pad\'e approximants to establish our notation. Given a function $f(s)$ analytic at $s=0$, we define its Pad\'e approximant as the ratio of two polynomials $R_N^J(s)$ and $S_N^J$ of degrees $N+J$ and $N$, respectively, such that its Taylor expansion at $s=0$ matches the one of $f(s)$ up to terms of order $s^{2N+J+1}$:
\beq
[N+J \; /\;N]_f(s)=\frac{R_N^J(s)}{S_N^J(s)} = f(s) + O(s^{2N+J+1}) \, .
\eeq
We assume that $S_N^J(0) \neq 0$. We say that the Pad\'e point is $s=0$.

The Pad\'e approximants of Stieltjes functions can be written in terms of orthogonal polynomials (see \cite{ourpade} for a concise derivation and definitions):
\beq
[N-1 \; / \;N]_f(s) = \frac{s^{-1} \rho_N(s^{-1})}{\pi_N(s^{-1})} .
\eeq
Here, $\pi_N$ and $\rho_N$ are, respectively, orthogonal polynomials and (first) associated orthogonal polynomials over the interval [0,1] with weight $W(u)$. The associated polynomials $\rho_N$ have degree $N-1$. This can be generalized to $J>-1$, and for our conformal propagators we find~\cite{ourpade}
\beq
\mathcal{P}_d(s-1)=[N+[\nu]\; /\; N]_{\Delta_d}(s-1) = \frac{\pi (\nu+1)}{\sin (\pi \nu)} \frac{s^J \hat{P}_N^{(\nu,[\nu]-\nu)}(\frac{2-s}{s})} {P_N^{(\nu,[\nu]-\nu)}(\frac{2-s}{s})}, 
\eeq
where $P_N^{(\alpha,\beta)}$ are Jacobi polynomials and $\hat{P}_N^{(\alpha,\beta)}$, their associated counterparts. In general one should add to this formula a polynomial in $s$, but it vanishes in our case thanks to the renormalization scheme we have chosen. It is important that we are using $J=[\nu]$ to ensure the correct UV behaviour. 
Alternatively, we can write
\begin{align}
\mathcal{P}_d(t) = & \Delta_d(t) - (-1)^{N}(\nu+1) \frac{\Gamma(N+\nu+1)\Gamma(n+[\nu]-\nu+1)}{\Gamma(2n+[\nu]+2)} \nn 
& \times \; \frac{(1+t)^{N+[\nu]+1} \; \mbox{}_2 F_1 \left(N+1,N+\nu+1,2N+[\nu]+2;\frac{t+1}{t}\right)} {t^{N+1} P_N^{(\nu,[\nu]-\nu)}\left(\frac{1-t}{1+t} \right)} .
\end{align}
In this form the rational structure is far from obvious, but it has the advantage of using only functions that are implemented in Mathematica.

\section{Unparticle vs particle signals}
The connection with orthogonal polynomials gives rise to nice properties of the Pad\'e approximants of unparticle propagators:
\begin{enumerate}
\item All the poles $t_i$ are simple, real and positive.
\item The residues are positive as well.
\item When $N \rightarrow \infty$ the Pad\'e approximant point-like converges to the exact propagator everywhere except along the positive real line. The rate of convergence is geometric for $|t+1|<1$.
\end{enumerate}
The first property follows simply from the facts that the zeros of orthogonal polynomials on a real interval are simple and belong to the interval, and that they are different from the zeros of the corresponding associated polynomials. The second one follows from the interlacing property of the zeros of orthogonal and associated orthogonal polynomials~\cite{szego}, which imply that the sign changes of the Pad\'e when a pole is crossed are neutralized by sign changes between poles, arising from the numerator. The proof of the third property is less straightforward and we do not give it here, but refer again the interested reader to Ref.~\cite{ourpade}. (See also~\cite{Baker} for algebraic proofs of these properties based on Pad\'e determinants). 

The first two properties are precisely what we need for a particle interpretation of the Pad\'e approximants. Indeed, for $1<d<2$, {\em $\mathcal{P}_d$ can be written as a sum of ordinary one-particle propagators with positive masses and residues.\/} When $d\geq 2$, on the other hand, we need to use higher degrees in the numerators and $\mathcal{P}_d$ is equal to a sum of one-particle propagators plus local terms. These local terms can always be put to zero by a change of renormalization scheme, which depends not only on the dimension but also on the particular Pad\'e approximant. At any rate, the local terms do not contribute to the spectral density.
The third property ensures a fast convergence in the Euclidean region, but not for timelike Minkowskian momentum. In fact, the spectral function of the Pad\'e approximants is a sum of delta functions. Nevertheless, the approximation can be good also for amplitudes in the Minkowskian region, if we take two effects into account. 

First, in real experiments the energy resolution of the detectors is finite. This means that the energy distribution of cross sections is convoluted with a function of the energy, representing the detector response with a resolution $\Delta \sqrt{t} \sim \Gamma(\sqrt{t})$. For small $\Gamma/\sqrt{t}$, this function should approach a delta function. Changing variables from $\sqrt{t}$ to $t$, we can approximately reproduce this effect by a convolution in $t$ of the spectral density with a distribution $\rho_\Lambda$, with an uncertainty $\Delta t \sim \Lambda(t) \sim \sqrt{t} \Gamma(\sqrt{t})$:
\beq
\sigma_{d,\Lambda}(t) = \int_0^\infty \sigma_d(w) \rho_\Lambda(w,t) \mathrm{d}w .
\eeq
Let us stick to $1<d<2$ and take
\beq
\rho_\Lambda(w,t)=\frac{1}{\pi} \frac{\Lambda}{(w-t)^2+\Lambda^2}.
\eeq
This is a representation of the delta function, so $\sigma_{d,\Lambda}\to \sigma_d$ when $\Lambda\to 0$. We want however to keep $\Lambda>0$ finite. Using
\beq
\rho_\Lambda(w,t) = - \frac{1}{\pi} \, \mathrm{Im} \frac{1}{t-w+i\Lambda}
\eeq
and analytically continuing \eref{spectral}, we find
\beq
\sigma_{d,\Lambda}(t) = -\frac{1}{\pi} \, \mathrm{Im} \Delta(t+i\Lambda) .
\eeq
Therefore, a finite resolution for unparticle (or particle) production can be implemented by evaluating the propagator at complex values\footnote{Observe that, in practice, it is not the energy of the unparticles which is measured, but rather the one of the jets and leptons produced together with them (or from their decays). The reconstruction of the (un)particles in the relevant processes, taking into account efficiencies, hadronization uncertainties and other effects, introduces a further smearing with a different energy dependence.  The encoding of these limitations into $\sigma_{d,\Lambda}$ is necessarily a rough approximation to effects that should be calculated on a process by process basis.\label{reconstruction}}.  These are good news for us, as the Pad\'e approximants converge uniformly if $t$ is kept off the positive real line. This sort of argument is standard in discussions of quark-hadron duality, and can be generalized to larger values of $d$.  We also note that a finite imaginary part in the argument of the propagator smooths down the behaviour of its real part too. However, in the virtual interference experiments the energy uncertainty is usually that of the beams, and thus smaller.

The second effect that smears the Pad\'e propagators is the fact that the coupling of the CFT to the SM gives corrections to the exactly conformal propagator. In particular, it generates an extra imaginary part, directly related to unparticle decay into SM particles~\cite{Rajaraman}. In our description, this means that we should add finite widths to the Pad\'e particles. This affects both the virtual exchange of unparticles and their production. Fortunately, this effect is again taken into account if we evaluate the propagator at momenta with a finite positive imaginary part.

To get a feeling of the goodness of the approximation, it is best to study explicit examples. For simplicity, we assume that $\Lambda$ is constant in the plots of the propagators below. In Fig.~\ref{fig_conformalprop} we plot the exact unparticle propagator with $d=4/3$ and the Pad\'e approximants $\mathcal{P}_{4/3}$ with $N=3$, $N=7$ and $N=15$, in terms of $t$ and with a fixed $\Lambda=0.1$. 
\FIGURE[t]{
\hspace{-1cm} \includegraphics[width=0.4\textwidth]{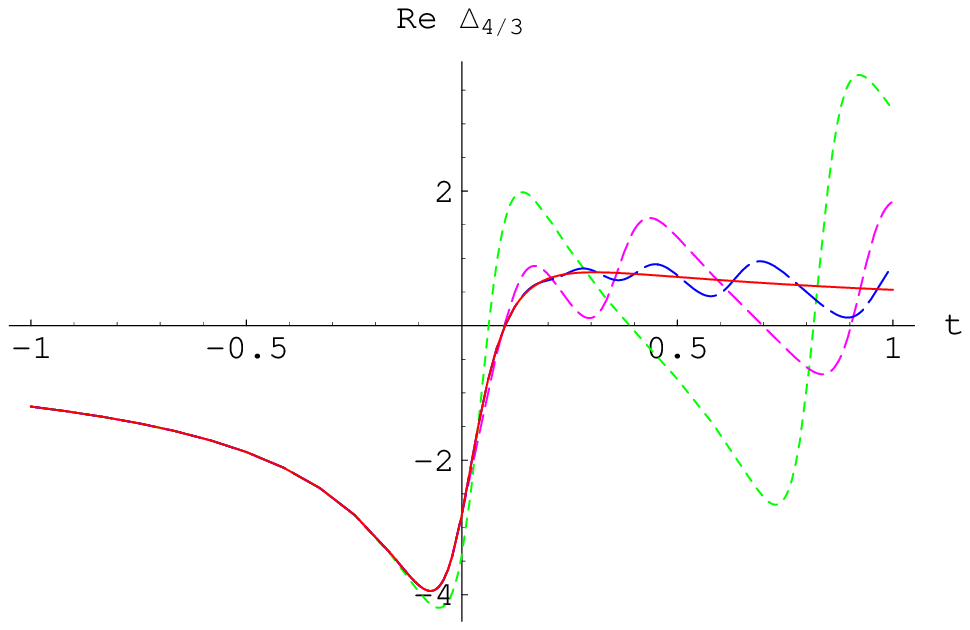}
\includegraphics[width=0.4\textwidth]{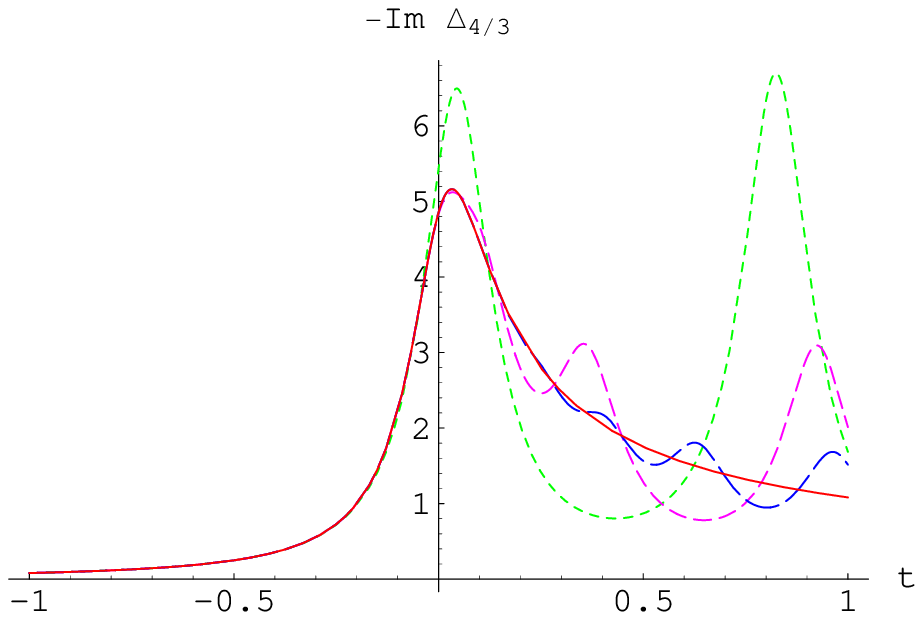}
\caption{{\footnotesize Real (left) and minus imaginary (right) parts of the unparticle propagator (red, solid) for $d=4/3$ and the corresponding Pad\'e approximants $\mathcal{P}_{4/3}$ with $N=3$ (green, short dashes), $N=7$ (purple, medium dashes) and $N=15$ (blue, long dashes). We take $\Lambda=0.1$. \label{fig_conformalprop} }}
}
The values of the squared masses (in units of $\mu^2$) and residues of the seven particles for $N=7$ are the following: 

\beq
\begin{array}{c|c}
\mathrm{Mass}^2 & \mathrm{Residue} \\
\hline 
0.0079221 & 0.352052 \\
0.107119 & 0.241402 \\
0.361388 & 0.241640 \\
0.924753 & 0.291822 \\
2.32400 & 0.431874 \\
7.21395 & 0.882060 \\
55.5609 & 4.26730 
\end{array}
\nonumber
\eeq
The fact that the number of masses above and below $\mu$ differ at most by 1 is general. 
We see in the plots that the approximation is very accurate in the Euclidean region, even for a number of particles as small as $N=3$. This would hold also for vanishing $\Lambda$. For Minkowskian timelike momentum, the Pad\'e propagator oscillates around the exact value. The accuracy in this region is better for small values of $t/\Lambda$. As expected, larger values of $t/\Lambda$ are well approximated when $N$ increases. We also see that, as one could have guessed, the approximation works well in the region where $\Lambda$ is greater than the mass spacings of the Pad\'e particles. The individual particles cannot be resolved in this region, so their masses and couplings cannot be determined. On the other hand, the convergence is slower for larger values of $d$.

The behaviour of the propagator translates directly into physical observables. As an instance, let us follow Ref.~\cite{G1} and consider the top decaying into an up quark and an unparticle via a derivative coupling
\beq
\lambda \bar{u} \gamma^\mu (1-\gamma_5) t \partial_\mu \mathcal{O}_d + \mathrm{h.c.}
\eeq
The normalized energy distribution of the outgoing up quark is displayed in Fig.~\ref{fig_decay} for unparticle stuff and for our particle approximation.
\FIGURE[t]{
\hspace{-1cm} \includegraphics[width=0.5\textwidth]{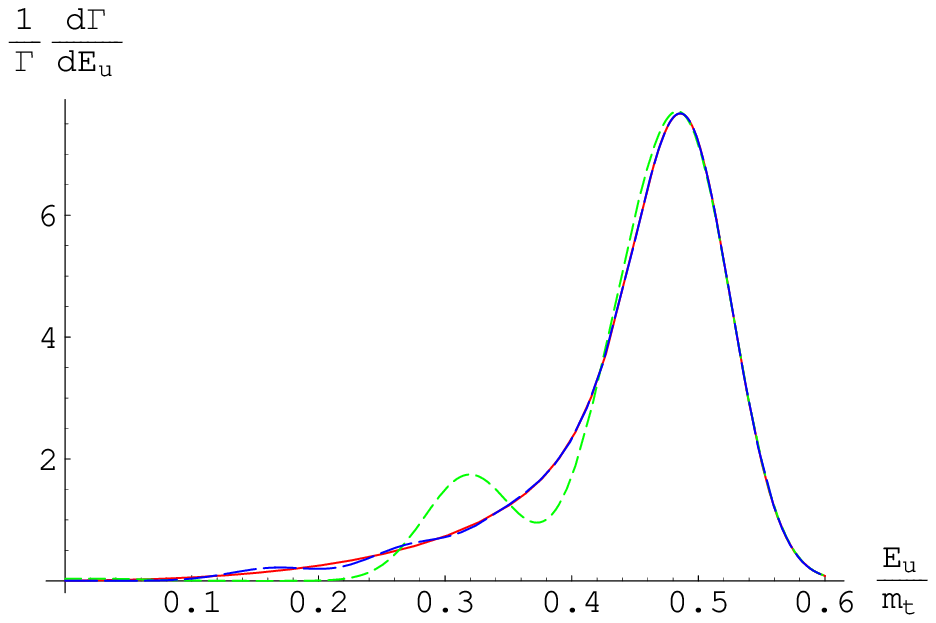}
\includegraphics[width=0.5\textwidth]{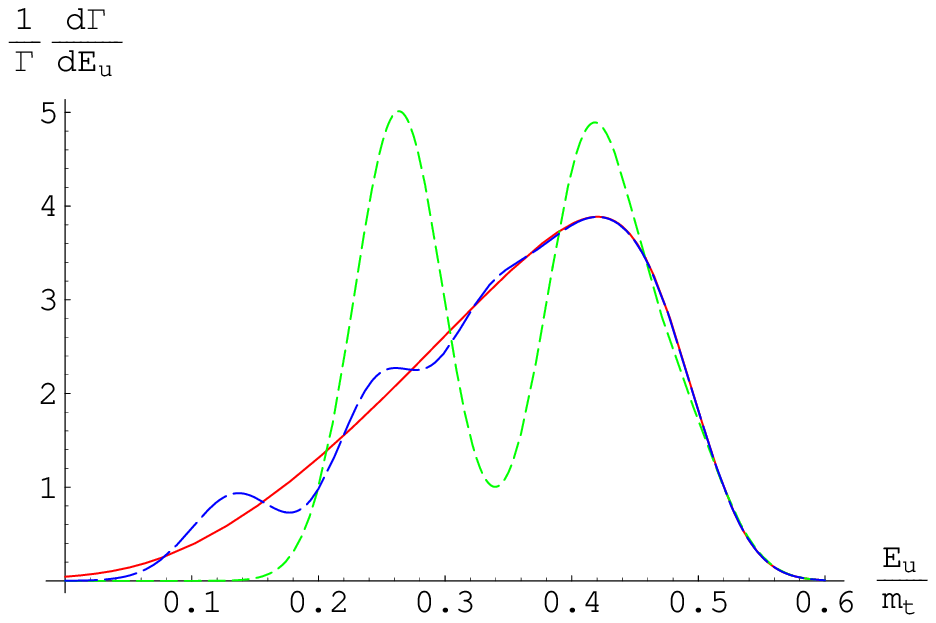}
\caption{{\footnotesize Normalized energy distribution of the up quark in $t\to u \mathcal{O}_d$ decay (red, solid) for $d=4/3$ (left) and $d=9/4$ (right), and the corresponding distributions for $t \to u \phi_i$ decay into Pad\'e particles $\phi_i$, with $N=7$ (green, short dashes) and $N=17$ (blue, long dashes). We choose the renormalization point $\mu=m_t$. The distributions are smeared with a gaussian of width $0.05 m_t$.
\label{fig_decay} }}
}
We consider the ideal situation in which the unparticle and the associated Pad\'e particles are stable and have no width. Finite widths would just make more difficult to distinguish both possibilities. 
This time, rather than going to complex energies, we have performed directly a gaussian smearing of the energy distribution of the up quark, using a constant width\footnote{The value of the width used for the plot, $\Gamma=0.05 m_t \approx 8.6$GeV, is somewhat above 5.3 GeV, the performance-goal resolution of the LHC hadronic calorimeters at energies around 85 GeV ~\cite{:2008zzm}, but is very optimistic when the other effects mentioned in the footnote $\mbox{}^{\ref{reconstruction}}$ are taken into account.}. The conformal dimension of the operator in the left and right plots are, respectively, $d=4/3$ and $d=9/8$. We see that the approximation is worse when the dimension increases. Observe that in this process, and for not too big dimensions, the approximation is much more accurate at the peak than in the tail on the left. Thus, in the presence of backgrounds it can be very difficult to distinguish the production of unparticles from the production of the Pad\'e particles, even for quite small $N$.

We finish this section with a generalization of our approach.
The reality properties of the Pad\'e approximants of Stieltjes functions do not hold any longer if we choose a complex Pad\'e point (instead of the real point $t=-1$). Even if the particle interpretation is lost in this case, it is interesting to observe that the approximation greatly improves in the physical region. We give an example of this in Fig.~\ref{fig.complex}. Notice that we are using a $\Lambda$  smaller than before. It is remarkable that just three {\em unphysical} particles are sufficient to give an almost perfect approximation to the unpartical propagator at any momentum.
\FIGURE[t]{
\hspace{-1cm} \includegraphics[width=0.4\textwidth]{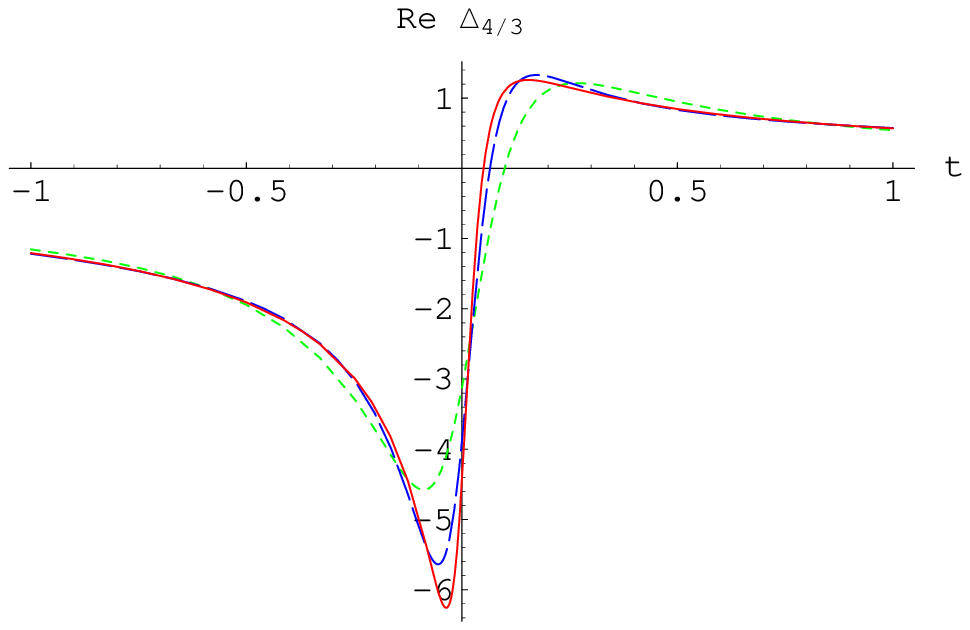}
\includegraphics[width=0.4\textwidth]{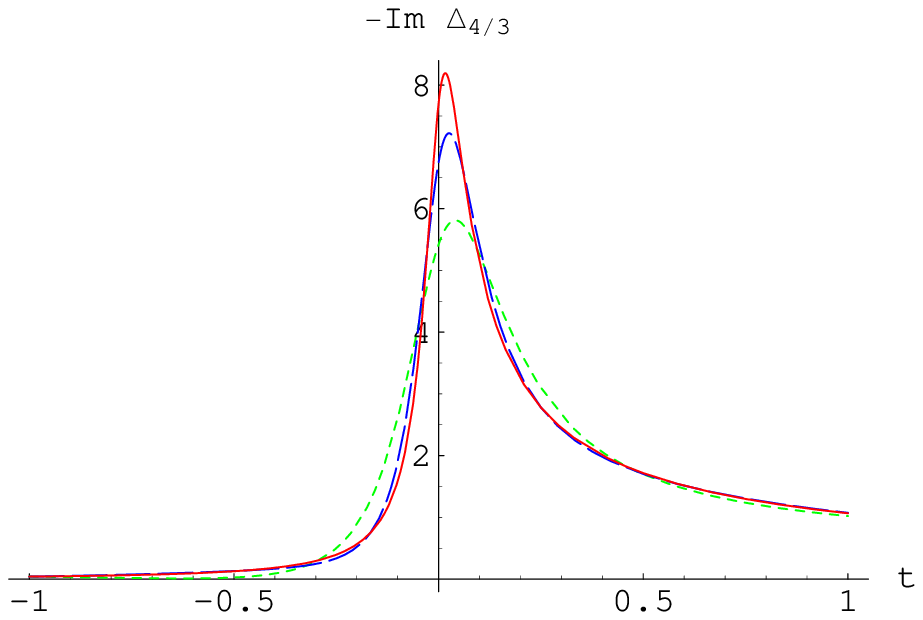}
\caption{{\footnotesize Real (left) and minus imaginary (right) parts of the unparticle propagator (red, solid) for $d=4/3$ and the corresponding Pad\'e approximants $\mathcal{P}_{4/3}$ with $N=2$ (green, short dashes) and $N=3$ (blue, long dashes). The Pad\'e point is in this case $t=i$, and we have chosen $\Lambda=0.05$.
\label{fig.complex} }}
}

\section{Unparticle properties from the particle perspective}
In this section we show that the Pad\'e approximants reproduce and help to understand some basic properties of the unparticle propagators. These properties are related to the behaviour of the propagators when $d$ approaches or crosses integer values. We carry out a numerical analysis, which should be sufficient to illustrate the general features.

Consider first the limit $d\to 1$. Taking N=6, $\mathrm{P}_{1.01}(t)$ has a very light particle with mass $m^2/\mu^2 = 0.00028$ and residue $0.968$, whereas the other five particles have residues smaller than $0.027$. So, we see explicitly how one particle in the Pad\'e smoothly becomes massless with residue 1 in the Pad\'e, while the other ones decouple. If we cross the unitarity limit, $d=1$, the original function is not Stieltjes any more, due to its IR behaviour. It can be seen explicitly that the light particle becomes tachyonic and the other ones get negative residues, small in absolute value if we stay near $d=1$. So, the Pad\'e approximants also reflect the pathologies of the original theory, and make explicit the ghosts below the unitarity bound. Moreover, it is interesting to see that this loss of unitarity comes about smoothly as we cross the limit $d=1$, and that it is small right below this limit.

Next, let us study what happens when we go from UV finite unparticles with $1<d<2$ to UV divergent unparticles with $2\leq d<3$. Consider the $[9/10]$ approximant $\mathcal{P}_{1.99}$. It has nine particles with masses below $5.13 \mu$, and one with $m^2/\mu^2=9950$ and residue $944146$. For energies of order $\mu$, this particle just provides a constant term equal to $94.9$. Let us compare this with the $[9/10]$ approximant to the unsubstracted propagator $\tilde{\Delta}_{2.01}$ (which is not Stieltjes). This has nine particles with masses and residues very similar to the previous ones. On the other hand, the heavy particle is tachyonic now, with $m^2/\mu^2=-10051$ and residue $1059380$. Hence, there is an instability, but if we are not far from $d=2$, it is located at large (spacelike) momenta. For small momentum, the tachyon behaves as a local constant term equal to -105.403. That the local term is that different from the previous one comes to no surprise, for the original unsubstracted propagator is also discontinuous as we cross $d=2$. In fact, it would be more appropriate to compare the substracted propagator with $d>2$, $\Delta_d$, to an oversubstracted propagator with $d<2$, $\bar{\Delta}_d(t)=\Delta_d(t)-\Delta_d(-1)$. These propagators are analytic continuations of each other. Of course, as we have seen in general before, the correct approach is to build the $[10/10]$ approximant to $\Delta_{2.01}$. This introduces an explicit constant term, $6.092$, and rearranges all the masses and residues to approximate the original function with non-tachyonic particles (with masses below $8.72 \mu$). Moreover, if we compare it with the $[10/10]$ approximant to $\bar{\Delta}_d$ with $d<2$, we do find a continuous evolution.

Summarizing, as we approach the limit in $d$ for a UV finite unparticle propagator, one particle of the $[N-1\;/\;N]$ Pad\'e approximant goes to infinity (without decoupling). After crossing this limit, it comes back from minus infinity to become a tachyon. This instability comes from the bad ultra-high-energy behaviour of the propagator (remember that we have performed the substraction at a finite renormalization point), and can be avoided introducing a constant term. This is done automatically by the $[N/N]$ Pad\'e. The same mechanism takes place every time an integer dimension $d\geq 3$ is crossed, when the correct Pad\'e $[N+[\nu]\; /\; N]$ is employed. On the other hand, if we insisted in using the $[N-1 \; / \; N]$ Pad\'e for $d \geq 3$, complex masses and residues would appear\footnote{This is closely related to the behaviour of Pad\'e approximants to non-Stieltjes functions, studied in~\cite{peris2} in the framework of large-N QCD.}.

\section{Unparticles with mass gap}
Everything so far can be generalized to unparticles with a mass gap. These arise when conformal invariance is broken in a very soft way that introduces a mass gap but preserves a continuum above it. For simplicity, we consider the particular propagator introduced in~\cite{Fox}:
\beq
\Delta_d^\mu(t)=\frac{\pi (\nu+1)}{\sin (\pi \nu)} (-t+\eta)^\nu + P_\nu(t),
\eeq
with $\eta=M^2/\mu^2$ and $M$ the mass gap. The spectral density is
\beq
\sigma_d^\eta(w) = (\nu+1)\, \theta(w-\eta) \, (w-\eta)^\nu.
\eeq
It is clear that all the previous results still hold in this case if we choose the Pad\'e point $t=\eta-1$. It follows that the approximations will be extremely accurate below the threshold for unparticle production. It is also possible in the presence of a mass gap to choose the Pad\'e point $t=0$. As pointed out in~\cite{peris1}, these Pad\'e approximants can be obtained in priciple from a chiral expansion of the theory. As an example, we plot in Fig.~\ref{fig.massgap} the unparticle propagator and its $[6/7]$ Pad\'e approximant, at point $t=0$, for $d=4/3$ and $\eta=1/3$. We choose $\Lambda=0.1$.
\FIGURE[ht]{
\hspace{-1cm}
\includegraphics[width=0.4\textwidth]{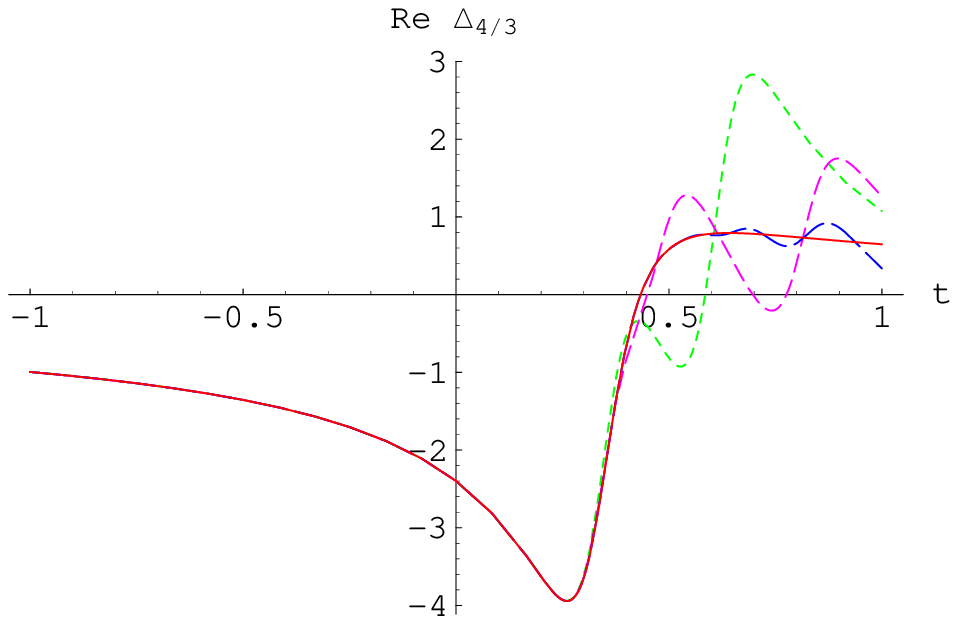}
\includegraphics[width=0.4\textwidth]{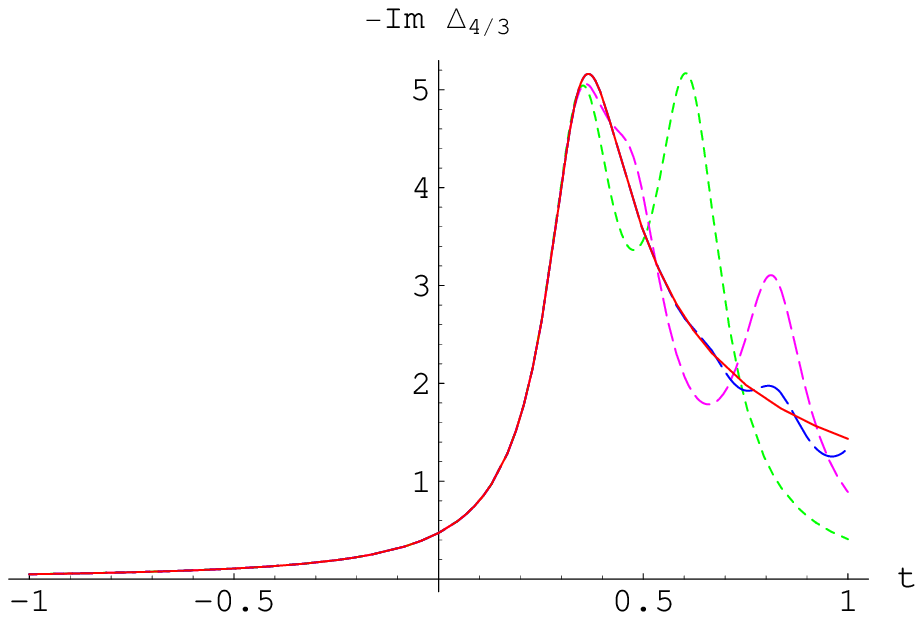}
\caption{{\footnotesize Real (left) and minus imaginary (right) parts of the unparticle propagator (red, solid) for $d=4/3$ and mass gap $\eta=1/3$, and the corresponding Pad\'e approximants $\mathcal{P}_{4/3}$ with $N=3$ (green, short dashes), $N=7$ (purple, medium dashes) and $N=15$ (blue, long dashes). The Pad\'e point is $t=0$, and $\Lambda=0.1$.
\label{fig.massgap} }}
}

\section{Conclusions}
We have shown that unparticle stuff with $1\leq d<2$ can be well approximated by a relatively small number of ordinary particles, with specific couplings and masses. This can be achieved in a systematic way by means of Pad\'e approximants. When $d \geq 2$, local terms are necessary in addition to the particle propagators. We conclude that a finite number of particles can give rise to the same production and interference signals as unparticles, once the limitations of experiments and the decay back into SM particles are taken into account\footnote{We do not claim that Pad\'e approximants give the best particle approximation to unparticle signals. For instance, tuning the widths of the individual particles, independently of the unparticle decay width, can improve the approximation.}. Intuitively, this is possible because these effects prevent the individual particles from being resolved. We should remark, nevertheless, that such particles have a very particular pattern of masses and couplings, with no {\it raison d'\^etre\/} without the principle of conformal invariance. It would be interesting to apply these ideas to specific processes including the SM backgrounds.

An alternative (exact) description of unparticles in terms of just one particle is known for $1\leq d<2$: a particle propagating in AdS space~\cite{Terning}. This follows from the AdS/CFT correspondence in the special case in which two alternative quantizations of the AdS free field exist~\cite{WK}. Unparticles with mass gap can be described by special (asymptotically AdS) geometries \cite{Terning,Gherghetta,infinite}. Similarly, the Pad\'e approximants in this paper can be given a higher-dimensional interpretation. Indeed, A. Falkowski and the author have shown in~\cite{ourpade} that, for Stieltjes functions, the Pad\'e numerators and denominators obey a second order difference equation. When $N\to \infty$, it becomes a field equation of motion in AdS\footnote{Migdal's double limit $N\to \infty$, $\mu \to \infty$ with fixed $\mu/N$ is instead dual to fields in a slice of AdS~\cite{Erlich,ourpade}. Here we leave $\mu$ fixed, as we do not want to arrive at a discrete spectrum.}. If we keep $N$ finite, the Pad\'e approximants are dual to particular deconstruction setups, described in~\cite{ourpade}. Therefore, from the holographic point of view, our approximation with a finite number of particles (KK modes) corresponds to both a discretization and a compactification of AdS space~\cite{Jorge}, in contrast to the compactification proposed in~\cite{Stephanov}.

\acknowledgments
I thank Juan Antonio Aguilar Saavedra, Adam Falkowski and Nuno Castro for useful comments. 
This work has been supported by MEC project FPA2006-05294 and
Junta de Andaluc{\'\i}a projects FQM 101, FQM 00437 and FQM03048.


\end{document}